\documentclass[aps,preprint,superscriptaddress]{revtex4-1}

\usepackage{amsmath}    % need for subequations
\usepackage{graphicx}   % need for figures
\usepackage{color}      % use if color is used in text
\usepackage{subfigure}  % use for side-by-side figures
\usepackage{hyperref}   % use for hypertext links, including those to external documents and URLs
\usepackage{todonotes}
\usepackage[latin1]{inputenc}

\begin{document}

\title{Dynamic response and electronic structure of potassium doped picene investigated by electron energy-loss spectroscopy}
\author{Friedrich Roth}
\author{Benjamin Mahns}
\author{Bernd B\"uchner}
\author{Martin Knupfer}
\affiliation{IFW Dresden, P.O. Box 270116, D-01171 Dresden, Germany}
\date{\today}

\begin{abstract}
We performed electron energy-loss spectroscopy studies in order to get a deeper insight into the electronic properties of potassium intercalated picene, a recently discovered superconductor. A comparison of the loss function of the undoped and doped compound shows the appearance of a new peak in the optical gap, which we attribute to the charge carrier plasmon. We find a dramatic increase for the value of the background dielectric constant $\epsilon_{\infty}$ upon doping. Our core level excitation data clearly signal filling of the conduction bands with electrons upon potassium addition.
\end{abstract}

\maketitle

\section{Introduction}
The discovery of a superconducting phase in the alkali metal doped fullerides \cite{Hebard1991,Gunnarsson2004} represents a breakthrough in the field of superconductivity, and attracted a lot of attention also on other molecular crystals built from $\pi$ conjugated molecules. In this context, further interesting phenomena were observed in alkali metal doped molecular materials such as the observation of an insulator-metal-insulator transition in alkali doped phthalocyanines \cite{Craciun2006}, a transition from a Luttinger to a Fermi liquid in potassium doped carbon nanotubes \cite{Rauf2004}, the formation of a Mott state in potassium intercalated pentacene \cite{Craciun2009}, or a potassium induced phase transition in iron phthalocyanine thin films.\cite{Roth2008} Also theoretical considerations led to fascinating predictions such as the formation of strongly correlated metals in the family of electron doped phthalocyanines.\cite{Tosatti2004} However, superconductivity with transition temperatures similar to those of the fullerides could not be observed in other molecular crystals until recently, when superconductivity has been reported for another alkali metal doped molecular solid, K-picene, with a transition temperature up to 18\,K.\cite{Mitsuhashi2010}

\par

Picene (C$_{22}$H$_{14}$) is a molecule that consists of five benzene rings arranged in a zigzag like manner and forms a herringbone, monoclinic crystal structure, similar to other aromatic molecules.\cite{Fraxedas2006} The lattice parameters are $a$ = 8.480\,\AA, $b$ = 6.154\,\AA, $c$ = 13.515\,\AA, and $\beta$ = 90.46$^\circ$ and the long molecular axis points towards the crystal $c$-direction.\cite{De1985}  Upon potassium addition, the crystal volume shrinks, but details of the crystal structure are unknown to date.

\par

Undoped picene is characterized by a large gap in the electronic spectrum (the optical onset is at about 3.2\,eV) and by four close lying conduction bands above the gap. These conduction bands are filled with electrons upon potassium addition. A recent photoemission study has demonstrated the appearance of a new spectral structure in the gap of picene as a function of K doping.\cite{Okazaki2010} In addition, a finite density of states at the Fermi levels has been reported.\cite{Okazaki2010} The electronic structure of K doped picene has also been addressed recently using calculations.\cite{Giovannetti2010,Kim2010,Andres2010,Kosugi2009} The results of these studies suggest that potassium doping to K$_3$picene leads to a partial filling of the conduction bands and the formation of a Fermi surface, i.\,e. a metallic ground state.

\par

In this contribution we report on an investigation of the electronic properties of potassium doped picene using electron energy-loss
spectroscopy (EELS). EELS studies of other undoped and doped molecular materials in the past have provided useful insight into their electronic properties.\cite{Kramberger2008,Schuster2007,Knupfer1999_2,Knupfer2001} We demonstrate that potassium addition leads to a filling of the conduction bands and the appearance of a low energy excitation which is attributed to the charge carrier plasmon of doped picene. In addition, our analysis of the spectra allows a determination of the dielectric function of K doped picene.

\section{Experimental}

\noindent Thin films of picene were prepared by thermal evaporation under high vacuum onto single crystalline KBr substrates kept at room
temperature with a deposition rate of 0.2\,nm/min and a evaporation temperature of about 440\,K. The film thickness was about 100\,nm. These picene films were floated off in destilled water,
mounted onto standard electron microscopy grids and transferred into the spectrometer. Prior to the EELS measurements the films were
characterized \textit{in-situ} using electron diffraction. All observed diffraction peaks were consistent with the crystal structure of
picene.\cite{De1985} Moreover, the diffraction spectra (cf. Fig.\, \ref{fig1} ) revealed a well pronounced texture, whereas the films show a
strong preference of crystallites with their $a,b$-plane parallel to the film surface. All electron diffraction studies and loss function
measurements were carried out using the 172\,keV spectrometer described in detail elsewhere.\cite{Fink1989} We note that at this high primary
beam energy only singlet excitations are possible. The energy and momentum resolution were chosen to be 85\,meV and 0.03\,\AA$^{-1}$,
respectively. We have measured the loss function Im[-1/$\epsilon(\textbf{q},\omega)$], which is proportional to the dynamic structure factor
S($\textbf{q},\omega$), for a momentum transfer $\textbf{q}$ parallel to the film surface [$\epsilon(\textbf{q},\omega)$ is the dielectric
function]. The C\,$1s$ and K\,$2p$ core level studies were measured with an energy resolution of about 200\,meV and a momentum resolution of
0.03\,\AA. In order to obtain a direction independent core level excitation information, we have determined the core level data for three
different momentum directions such that the sum of these spectra represent an averaged polycrystalline sample also for the oriented or textured
samples.\cite{Egerton1996} The core excitation spectra have been corrected for a linear background, which has been determined by a linear fit of
the data 10\,eV below the excitation threshold.

Potassium was added in several steps by evaporation from commercial SAES (SAES GETTERS S.p.A., Italy) getter source under ultra-high vacuum
conditions (base pressure lower than 10$^{-10}$\,mbar) until a doping level of about K$_3$picene was achieved. In detail, in each doping step, the
sample was exposed to potassium for 5\,min, the current through the SAES getter surce was 6\,A and the distance to the sample was about 30\,mm.
During potassium addition, the film was kept at room temperature. Post-annealing of the films at 440\,K for several hours did not lead to
changes in the doping level, which demonstrates that potassium diffusion at room temperature is sufficient to achieve a homogenous distribution
in the film.

In order to perfom a Kramers-Kronig analysis (KKA), the raw date have been corrected by substracting contributions of multiple scattering
processes and elimination of contributions of the direct beam by fitting the plasmon peak with a model function.\cite{Sing1999}

\section{Results and discussion}

\begin{figure}[h]
\includegraphics[width=0.49\textwidth]{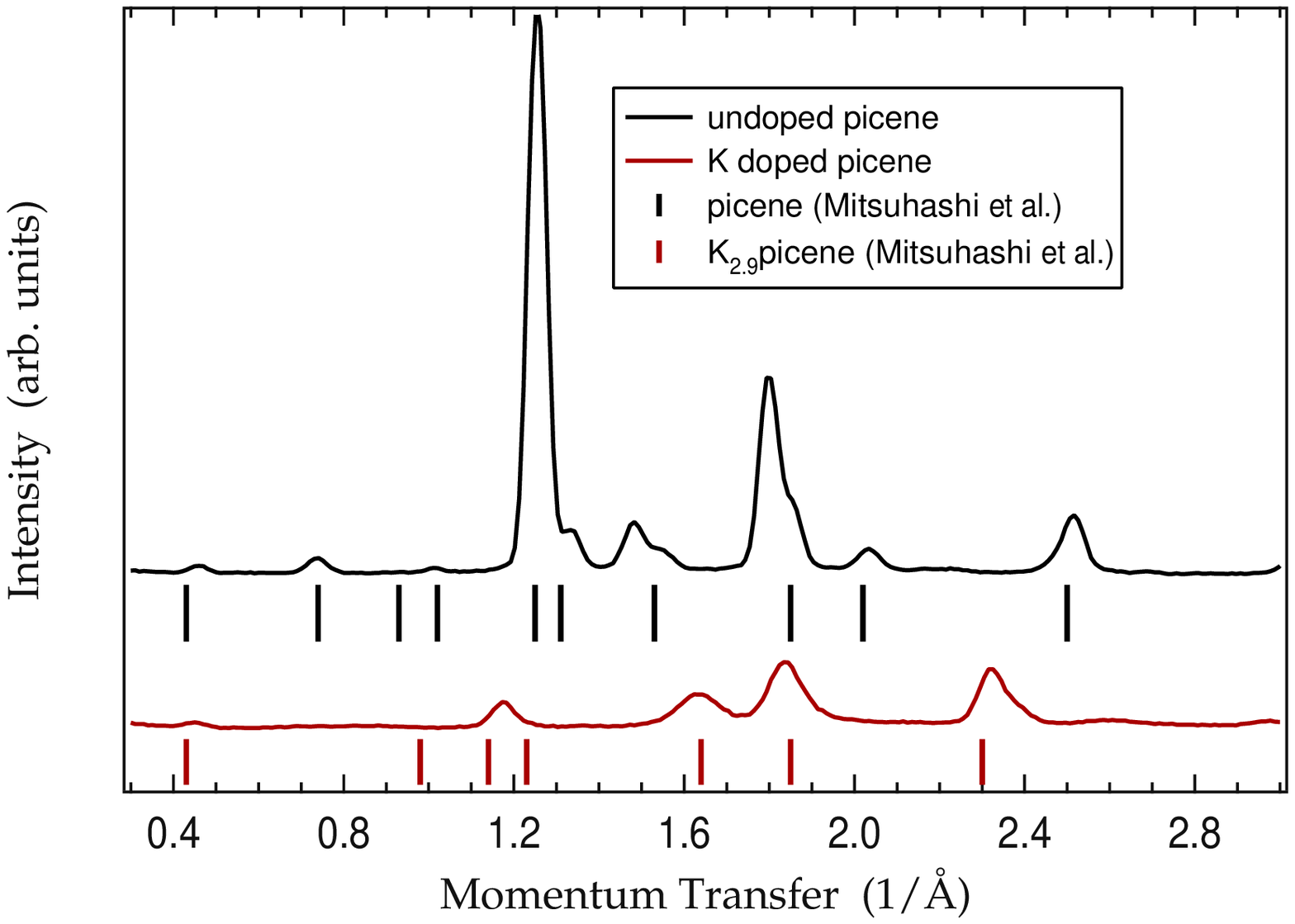}
\includegraphics[width=0.49\textwidth]{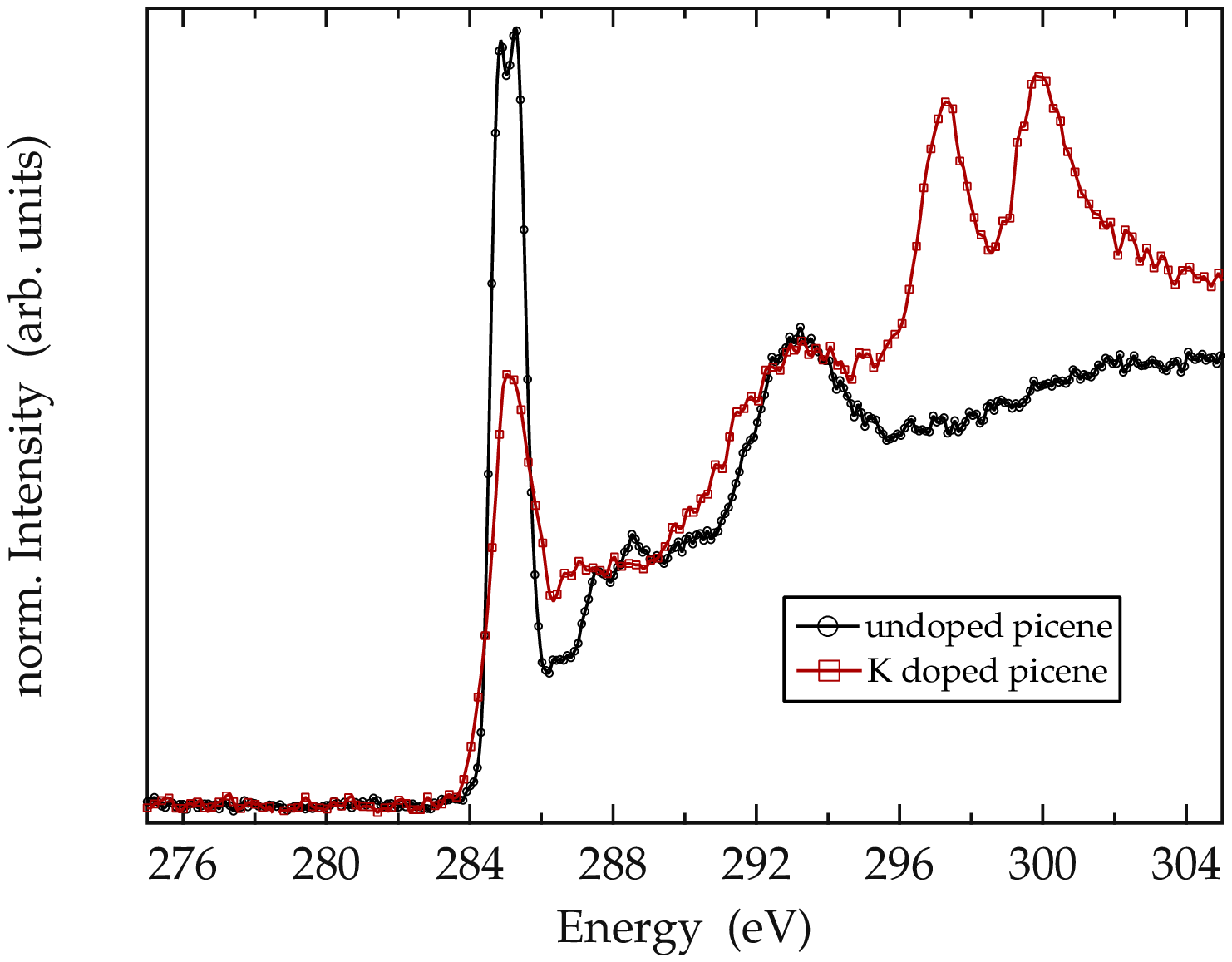}
\caption{\label{fig1}Left panel: Comparison of the elastic diffraction profiles of solid picene for the undoped (black/upper curve) and
potassium doped (red/lower curve) case. The ticks below the curves indicate the positions of the x-ray diffraction peaks as reported in
\cite{Mitsuhashi2010}. Right panel: C\,$1s$ and K\,$2p$ core level excitations of undoped and potassium doped picene.}
\end{figure}

\noindent In the left panel of Fig.\,\ref{fig1} we show representative electron diffraction profiles of undoped and potassium doped picene
films. An analysis of the diffraction peaks of undoped picene with calculated positions based on the crystal parameters reveals that our films
consist mostly of crystallites with their $a,b$-plane parallel to the film surface. The diffraction profile can be reproduced well with the
X-ray diffraction pattern of pristine picene as reported in \cite{Mitsuhashi2010} as seen as black ticks below the curve (note that the X-ray
diffraction pattern represents all three directions in contrast to our $a,b$-textured films). Upon potassium doping the diffraction profile
drastically changes. The diffraction peaks at about 1.25, 1.48, 2.05 and 2.52\,\AA$^{-1}$ disappear, instead some new peaks around 1.18, 1.63
and 2.31\,\AA$^{-1}$ are observed. This observation is in very good agreement with the published X-ray diffraction pattern for K$_{2.9}$picene.\cite{Mitsuhashi2010}
In our studies, further doping does not change the elastic scattering profile substantially, which indicates that with our preparation procedure
a doped phase with a stoichiometry close to K$_{3}$picene is formed. The addition of more potassium results in the formation of a metallic K overlayer on our films.

\par

In order to be able to further analyze the doping induced changes, we have additionally measured the C\,$1s$ and K\,$2p$ core excitation edges.
These data can be used to determine the stoichiometry of the potassium doped picene films. Moreover, the C\,$1s$ excitations
represent transitions into empty C $2p$-derived levels, and thus allow to probe the projected unoccupied electronic density of states of
carbon-based materials.\cite{Soininen2005,Knupfer1999,Knupfer1995} Fig.\,\ref{fig1} (right panel) shows the C\,$1s$ and K\,$2p$ core excitations
for potassium doped picene in comparison to that of pure picene. Both spectra were normalized in the region between 291\,eV and 293\,eV, i.\,e. to
the $\sigma^*$ derived intensity. For the undoped case, we can clearly identify a sharp and strong feature in the range between 284 - 286\,eV,
and some additional broad features at 289\,eV and 293\,eV. Below $\sim$ 290\,eV the structures can be assigned to transitions into $\pi^*$
states representing the unoccupied electronic states. Recent high resolution data have additionally documented a fine structure of the low
lying, predominant feature due to the four close lying conduction bands above the energy gap of picene.\cite{Roth2010} Above 290\,eV the
spectral weight is dominated by C $1s-\sigma^*$ transitions. Also in the case of potassium doped picene the data still reveal some fine
structure which demonstrates the molecular derived nature of the electronic states. Again, the spectrum is dominated by a sharp excitation
feature right after the excitation onset at 284\,eV and, in addition, by K $2p$ core excitations, which can be observed at 297.2\,eV and
299.8\,eV, and which can be seen as a first evidence of the successful doping of the sample. The step-like structure at about 287\,eV, which
can be clearly seen in the undoped case, is not observed. Importantly, a clear reduction of the spectral weight of the first C\,$1s$ excitation
feature by a factor of about 0.62  is observed in Fig.\,\ref{fig1} upon doping. Taking into account the four conduction bands that contribute
to this intensity in the undoped case, this reduction is a clear signal for the successful doping and it can be used to analyze the doping (filling) level of the conduction bands. We thus arrive
at a filling of these four levels by 3 electrons, in very good agreement to our stoichiometry analysis of K$_{3}$picene and the fact that
each potassium provides its outer $s$ electron for the doping process.

\par

The stoichiometry analysis can further be substantiated by comparing the K\,$2p$ and C\,$1s$ core excitation intensities in comparison to other doped molecular films with known stoichiometry, such as K$_6$C$_{60}$ \cite{Knupfer2001} or K$_4$CuPc \cite{Flatz2007}. Details of this procedure can be found in previous publications.\cite{Flatz2007,Roth2008} The results shown in the right panel of Fig.\,\ref{fig1} indicate a doping level of K$_{2.8}$picene, which again is in very good agreement to the other results discussed above.

\begin{figure}[t]
\includegraphics[width=0.6\textwidth]{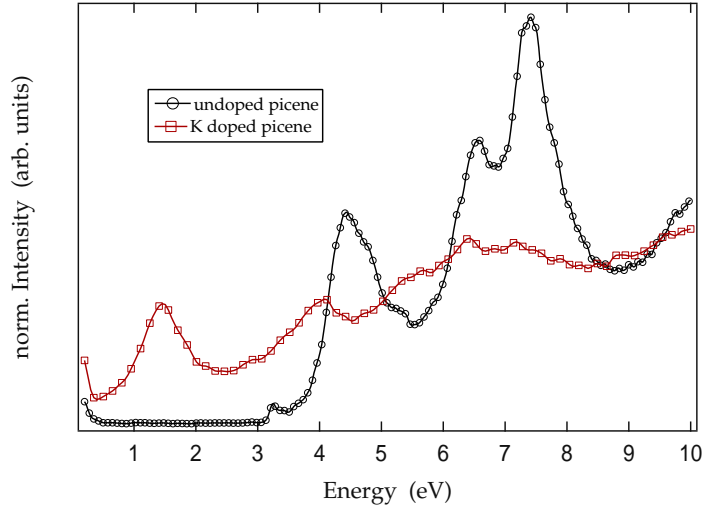}
\caption{\label{fig2}Comparison of the loss function in the range of 0 - 10\,eV for undoped (black circles) and K intercalated picene (red
squares).}
\end{figure}

Doping of picene also causes major changes in the electronic excitation spectrum as revealed in Fig.\,\ref{fig2}, where we show a comparison of
the loss functions in an energy range of 0-10\,eV measured using EELS. These data are taken with a small momentum transfer $\textbf{q}$ of
0.1\AA$^{-1}$, which represents the optical limit. For undoped picene, we can clearly identify maxima at about 3.3\,eV, 4.6\,eV, 6.4\,eV and
7.3\,eV, which are due to excitations between the occupied and unoccupied electronic levels and which recently have been well reproduced using
density functional based calculations.\cite{Roth2010} Note that the spectra in Fig.\,\ref{fig2} represent predominantly excitations with a
polarization in the $a,b$ crystal plane. Upon doping, the spectral features become broader, and a downshift of the major excitations can be
observed. We assign this downshift to a relaxation of the molecular structure of picene as a consequence of the filling of anti-bonding $\pi^*$
levels. The reason for the spectral broadening is unclear at present. We can only speculate that it might be connected 
to a reduced life time of the electronic excitation in the metallic state and/or to the formation of a doped sample with very small grain sizes. In addition, for the doped film a new structure at about 1.5\,eV is observed in the former gap of picene.

\par

\begin{figure}[t]
\includegraphics[width=0.6\textwidth]{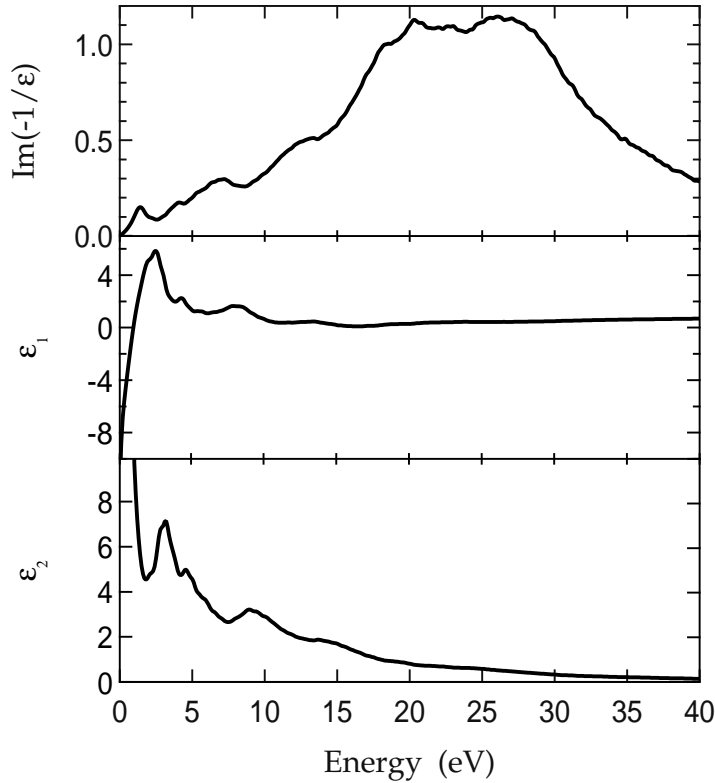}
\caption{\label{fig3}Loss function (Im(-1/$\epsilon$), real part ($\epsilon_1$) and imaginary part ($\epsilon_2$) of the dielectric function of K doped picene. The momentum transfer is $q$\,=\,0.1\,\AA$^{-1}$. Note that in contrast to Fig.\,\ref{fig2}, the loss function is corrected for the contribution of the direct beam.}
\end{figure}

In order to obtain deeper insight into these variations, we have analyzed the measured loss function, Im(-1/$\epsilon$), of doped picene using a
Kramers Kronig analysis.\cite{Fink1989} Such a KKA analysis provides us with the dielectric function, and in Fig.\,\ref{fig3} we present the
results of this analysis in a wide energy range. As the dielectric function, and also the loss function, are so-called response functions (and therefore
so-called retarded Green's functions) they reveal several very useful properties. In particular they satisfy certain sum rules, which are
important to test the consistency of our KKA analysis. One of these sum rules relates the loss function and the dielectric function to the
density, $N$, of all valence electrons:

\begin{align*}
 \int\limits_{0}^{\infty} d\omega \omega \operatorname{Im}\left(-\frac{1}{\epsilon(\textbf{q},\omega)}\right) = \int\limits_{0}^{\infty} d\omega \omega \epsilon_2 = \omega_p^2 \cdot \frac{\pi}{2} \propto
 N
\end{align*}

An evaluation of this sum rule for the loss function and the dielectric function after our KKA results in a very good agreement of the two
values with an error of less than 1.5\,\%. Furthermore, these values are also in good correspondence to what is expected from a calculation of
the electron density of doped picene, we find a deviation of 7\,\% only.

For metallic systems a further sum rule can be employed \cite{Mahan2000}, which allows an additional check of our analysis:

\begin{align*}
 \int\limits_{0}^{\infty} d\omega \frac{\operatorname{Im}\left(-\frac{1}{\epsilon(\textbf{q},\omega)}\right)}{\omega} = \frac{\pi}{2}.
\end{align*}

Here, after our KKA we arrive at a value of 1.55, very close to the expectation of $\frac{\pi}{2}$.

In Fig.\,\ref{fig3}, the loss function (upper panel) is dominated by a broad maximum in the range between 20 - 25\,eV which can be assigned to
the $\pi + \sigma$ plasmon, a collective excitation of all valence electrons in the system. Its energy position is similar to that observed for
other $\pi$ conjugated materials.\cite{Fink1989,Knupfer2001} Various interband excitations at 2 - 20\,eV can be observed as maxima in the imaginary part of the
dielectric function, $\epsilon_2$ (lower panel). Most interestingly, the new low energy excitation at about 1.5\,eV in the loss function is not
represented by a maximum in $\epsilon_2$ but by a zero crossing near 1.4\,eV in the real part of the dielectric function, $\epsilon_1$.
Therefore, this spectral feature represents a collective excitation (density oscillation), and we assign it to the charge carrier plasmon of
doped picene.

\begin{figure}[t]
\includegraphics[width=0.6\textwidth]{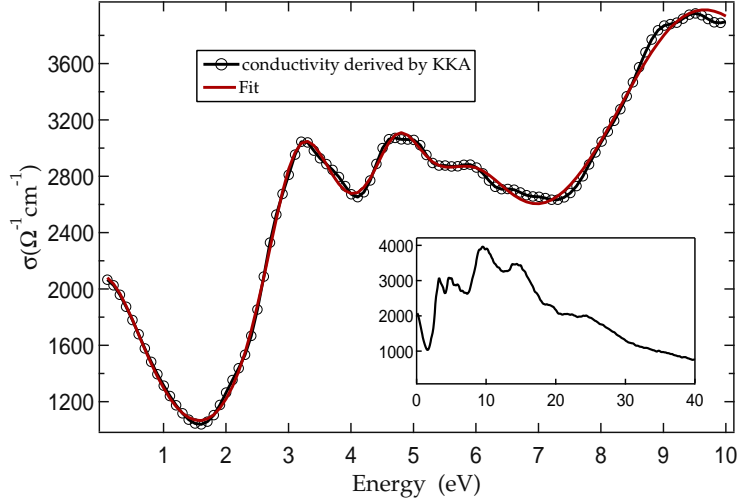}
\caption{\label{fig4}Optical conductivity $\sigma = \omega\epsilon_0\epsilon_2$ of K doped picene for a momentum transfer of
$q$\,=\,0.1\,\AA$^{-1}$ (black circles) derived by a Kramers-Kronig analysis of the loss function. Additionally, the result of a Drude-Lorentz
fit is shown (red line). The inset shows the optical conductivity in a larger range up to 40\,eV.}
\end{figure}

To elucidate all the observations further, we show in Fig.\,\ref{fig4} the optical conductivity, $\sigma$, of doped picene. The optical
conductivity is directly proportional to the matrix element weighted joint density of states of the corresponding excitation. As revealed by
Fig.\,\ref{fig4}, $\sigma$ consists of a free electron contribution at low energies due to intraband transitions in the conduction bands and
some additional interband contributions. In order to obtain a more detailed picture and to separate intra- and interband contributions, we have
fitted the optical conductivity in the range between 0 - 30\,eV using a simple Drude-Lorentz model:

\begin{align}
\epsilon(\omega) = 1 - \frac{\omega_D^2}{\omega^2+i\gamma_D\omega} + \sum_{j} \frac{f_j}{\omega_{j_o}^2-\omega^2-i\gamma_j\omega}
\end{align}

The Drude (charge carrier) part ($\omega_D$,$\gamma_D$) describes the free electron contribution and the Lorentz oscillators ($f_j$, $\gamma_j$
and $\omega_{j_o}$) represent the interband transitions. The resulting fit parameters are given in Table\,\ref{tab1} and are also shown in
Fig.\,\ref{fig4}. This Figure demonstrates that our model description of the data is very good. We note, that the result of our fit also
describes the real part of the dielectric function very well, which demonstrates the consistency of our description.

\begin{table}[h]
  \centering
\begin{tabular}{c| c c c || c c}
    \hline
  i \quad & \quad $\omega_{j_o}$ (eV) & \quad  $\gamma_j$ (eV) & \quad  $f_j$ (eV) &\quad  $\gamma_D$ (eV) & \quad  $\omega_D$ (eV)\\
  \hline
  1 \quad &\quad  2.05 & \quad 0.15 & \quad  0.31 &\quad  1.14 & \quad  4.22\\
  2 \quad &\quad  3.22 & \quad 1.83 & \quad  5.74 & & \\
  3 \quad &\quad 4.77 & \quad  1.44 & \quad  3.79 & & \\
  4 \quad &\quad  6.01 & \quad  2.16 & \quad  4.34 & & \\
  5 \quad &\quad  9.53 & \quad  4.94 & \quad  10.62 & & \\
  6 \quad &\quad  14.75 & \quad  6.85 & \quad  10.49 & & \\
  7 \quad &\quad  24.53 & \quad  16.39 & \quad  13.09 & & \\
\hline
 \end{tabular}
\caption{Parameters derived from a Drude-Lorentz fit of the optical conductivity (as shown in Fig.\,\ref{fig4}) using formula (1). The Drude
part is given by the plasma energy $\omega_D$ and the width of the plasma (damping) $\gamma_D$, while $f_j$, $\gamma_j$ and $\omega_{j_o}$ are
the oscillator strengh, the width and the energy position of the Lorentz oscillators.} \label{tab1}
\end{table}

We arrive at an unscreened plasma frequency $\omega_D$ of about 4.2\,eV, which now can be compared to the expectation value for this plasma
frequency taking into account the six conduction electrons per unit cell (2 picene molecules per unit cell) in K$_3$picene:

\begin{align*}
 \hbar\omega_D = \sqrt{\frac{ne^2}{\epsilon_{0}m_e}} = 3.52\,\mathrm{eV}
\end{align*}

This value is somewhat lower than what we have derived using our fit procedure. This might be related to an effective mass of the charge
carriers in doped picene, which is reduced as compared to the free electron value, $m_0$. For related (undoped) organic crystals such as rubrene
\cite{Machida2010}, PTCDA \cite{Temirov2006,Ueno2008} or pentacene \cite{Doi2005} an effective mass also lower than $m_0$ has been deduced. In the
case of superconducting, K doped picene, future theoretical and experimental studies are necessary to clarify this point.

\par

Finally, the difference between the (unscreened) plasma frequency as obtained in our fit and the observed (screened) value that can be read off the loss function (about 1.5\,eV) or the zero crossing of the real part of the dielectric function (about 1.4\,eV), can be described by a background dielectric constant $\epsilon_{\infty}$, which effectively describes the screening of the plasma oscillations due to all higher lying electronic excitation of the system. This results in:

\begin{align*}
 \epsilon_{\infty} \sim \left(\frac{4.22}{1.45}\right)^2 \sim 8.5
\end{align*}

Surprisingly, this value is significantly larger than the static dielectric constant of undoped picene, where we derived a value of 3 - 3.5 (which is in a very good agreement with the calculated value of $\epsilon_1(E=0)$ \cite{Roth2010}). In other words, the screening ability of doped picene seems to be much larger than that of the parent compound, whereas such a dramatic change in the background dielectric constant is for instance not observed going from undoped to doped C$_{60}$.\cite{Gunnarsson1996,Iwasa1992,Degiorgi1994} It is tempting to relate this increased screening ability to the appearance of superconductivity, since the pairing of charge carriers has to overcome their Coulomb repulsion, but this issue certainly needs further investigations from both, experiment and theory to be settled.

\section{summary}

To summarize, we have investigated the electronic properties of potassium doped picene using electron energy-loss spectroscopy. 
Electron diffraction and core level excitation data signal the formation of a doped phase with a stoichiometry close to K$_3$picene. 
The reduction of the lowest lying C\,$1s$ excitation features clearly demonstrates that potassium addition leads to a filling of the picene conduction bands. The electronic excitation spectrum changes substantially upon doping. In particular, a new low energy feature is observed at about 1.5\,eV, which is assigned to the charge carrier plasmon of doped picene. Carrying out a Kramer Kronig analysis we find that the background dielectric screening is significantly larger for doped picene as compared to the parent compound, which might be related to the appearance of superconductivity in doped picene.

\begin{acknowledgments}
We thank R. Sch\"onfelder, R. H\"ubel and S. Leger for technical assistance. This work has been supported by the Deutsche Forschungsgemeinschaft (grant number KN393/12 and KN393/14).
\end{acknowledgments}

\bibliography{Pic_doping}
\bibliographystyle{apsrev4-1}
\end{document}